\documentclass[preprint2]{aastex}

\shorttitle{Atmospherically Induced Ellipticities}
\shortauthors{De Vries et al.}

\newcommand{\ro}{r$_{\rm o}$\,}

\begin{document}

\title{Image Ellipticity from Atmospheric Aberrations}

\author{W. H. de Vries\altaffilmark{1,2}, S. S. Olivier\altaffilmark{3}, S. J. Asztalos\altaffilmark{3}, L. J. Rosenberg\altaffilmark{3,4}, and K. L. Baker\altaffilmark{3}}

\altaffiltext{1}{University of California, Department of Physics, 1 Shields Ave, Davis, CA 95616}
\altaffiltext{2}{Institute for Geophysics and Planetary Physics, LLNL, L-413, 7000 East Avenue, Livermore, CA 94550}
\altaffiltext{3}{Lawrence Livermore National Laboratory, 7000 East Avenue, Livermore, CA 94550}
\altaffiltext{4}{University of Washington, Department of Physics, Box 35160, Seattle, WA 98195}
\email{devries1@llnl.gov}

\begin{abstract}

We investigate the ellipticity of the point-spread function (PSF)
produced by imaging an unresolved source with a telescope, subject to
the effects of atmospheric turbulence. It is important to quantify
these effects in order to understand the errors in shape measurements
of astronomical objects, such as those used to study weak
gravitational lensing of field galaxies.  The PSF modeling involves
either a Fourier transform of the phase information in the pupil plane
or a ray-tracing approach, which has the advantage of requiring fewer
computations than the Fourier transform. Using a standard method,
involving the Gaussian weighted second moments of intensity, we then
calculate the ellipticity of the PSF patterns. We find significant
ellipticity for the instantaneous patterns (up to more than
10\%). Longer exposures, which we approximate by combining multiple
($N$) images from uncorrelated atmospheric realizations, yield
progressively lower ellipticity (as $1/\sqrt{N}$).  We also verify
that the measured ellipticity does not depend on the sampling interval
in the pupil plane using the Fourier method.  However, we find that
the results using the ray-tracing technique do depend on the pupil
sampling interval, representing a gradual breakdown of the geometric
approximation at high spatial frequencies.  Therefore, ray tracing is
generally not an accurate method of modeling PSF ellipticity induced
by atmospheric turbulence unless some additional procedure is
implemented to correctly account for the effects of high spatial
frequency aberrations.  The Fourier method, however, can be used
directly to accurately model PSF ellipticity, which can give insights
into errors in the statistics of field galaxy shapes used in studies
of weak gravitational lensing.

\end{abstract}


\section{Introduction}

Statistical analyses of weak gravitational lensing of field galaxies
\citep[e.g.,][]{wittman00,vanwaerbeke00,bacon00} are being used as
probes of cosmology and are expected to provide some of the strongest
cosmological tests in future, large astronomical survey projects, such
as the Large Synoptic Survey Telescope \citep[LSST;
e.g.,][]{tyson01,tyson02}. These surveys will allow for the precise
determination of various cosmological parameters, such as the matter
density distribution $\Omega_m$, the cosmological constant
$\Omega_\Lambda$, the equation of state $w$ of the dark energy, and
its time derivative. This is done by accurately analyzing large
numbers of background galaxies as their shapes are sheared by
intervening large-scale structure through weak gravitational lensing,
the results of which are then combined with, e.g., the very accurate
measurements of the cosmic microwave background radiation by the WMAP
satellite.

A critical part in these analyses is the accuracy to which one can
measure and correct the shape of the Point Spread Function (PSF) as it
varies across the detector \citep[see, e.g.,][]{hoekstra04}. This PSF
anisotropy is largely induced by the atmosphere
\citep[e.g.,][]{wittman05}, and cannot easily be modeled without
incorporating an explicit atmosphere. It is possible to mimic the
effects of the atmosphere by convolving either artificially generated,
or high resolution HST images with a suitable PSF (see, e.g., Heymans
et al. 2006a for the former, and Bacon et al.  2001 for the latter
approach), but that still does not include effects of PSF
anisotropy. We therefore set out to model the behavior of the PSF as
it gets folded through a realistic atmosphere and telescope
system. Since ray-tracing methods are commonly used to simulate the
shearing signal of weak lensing by large scale structure
\citep[e.g.,][]{jain00,vale03,heymans06b}, we include both ray-tracing
and Fourier transform methods to calculate what the PSF should be
based on the phase and intensity information in the pupil plane of the
telescope. The ray-tracing method has the advantage of computational
speed compared with the Fourier transform method. However, the Fourier
transform method correctly treats the effects of interference, which
are ignored by the geometric ray-tracing method. Because of its
relative computational efficiency, it is useful to understand whether
the ray tracing approach gives adequate PSF ellipticity information.
More importantly, it is crucial to understand the elliptical
properties of the PSF in order to enable quantitative analysis of the
errors in statistical studies of field galaxy shapes.

Our modeled observational setup is characterized as follows: a generic
8 m-class telescope, a turbulent atmosphere with a Kolmogorov power
spectrum, and a single, on-axis point-source located at infinity. The
ellipticity and its direction are assumed to be representative for a
single region over which these quantities do not vary. The angular
extent of these regions can be, depending on observing conditions,
larger than an arcminute \citep[e.g.,][]{asztalos06}. Since the
typical LSST exposure will be 15 seconds, we also investigate the time
dependence of the atmospherically induced ellipticities, and how it
imposes limits on the ability to measure them.

\subsection{Layout of the paper}

The paper is organized as follows. In \S~\ref{simmethod} and
\S~\ref{calc}, we discuss the atmospheric simulations and the methods
applied to calculate the PSF ellipticity. Then in \S~\ref{results}, we
describe the results for both the ray-tracing method (RTM) and the
Fourier transform method (FTM). The main difference between these two
is the way one models image formation by the telescope. The RTM traces
the geometric path of rays from the pupil plane onto the focal plane,
whereas the FTM applies a Fourier transform to the complex
electromagnetic field in the pupil plane in order to calculate the
resulting image. The latter correctly incorporates effects of
interference, unlike the RTM. The geometric approximation of the RTM
has implications for shape measurements in the image plane, which we
quantify in this paper.

The results are subdivided into the effects of pupil plane sampling
(\S~\ref{sampling}), the variation of ellipticity as a function of
exposure time and the presence of wind (\S~\ref{wind}), and seeing
(\S~\ref{seeingEff}). The latter is approximated by using varying
ratios of D / \ro (where D is the aperture size of the telescope, and
\ro is the coherence length of the atmosphere). Longer exposure times
are simulated by increasing the number of completely independent phase
screens (which is a function of aperture diameter and wind-speed). In
order to check the accuracy of this approximation, we also evaluate a
model that includes intermediate phase screens, i.e., screens that
are not completely decorrelated from the previous one, but are
translated by a small fraction of the aperture size along the wind
direction.  In each of these sections we investigate the differences
between the RTM and FTM, which are summarized in \S~\ref{conclusion}.

\section{Simulations and Methods}\label{simmethod}

We generated random phases for the electromagnetic field in the pupil
of the telescope using Kolmogorov statistics to represent the effects
of atmospheric turbulence. The inner turbulence scale of the
simulations is set by the pixel size used in the simulations, to a
fraction of \ro. The outer scale of the turbulence has been fixed to a
value larger than the simulation box ($> 800$ m for the large screens
in \S~\ref{wind}, for instance), so that effectively the simulations
see a Kolmogorov turbulence spectrum with an infinite outer scale.
Assuming a constant magnitude in the circular pupil, we propagated the
field to the focal plane using a Fourier transform, and by squaring
the resulting magnitude, created a representation of the focal plane
image intensity. These images contain distinct speckle patterns for
different realizations of the atmosphere.

The ray tracing algorithm used an idealized model of a reflecting
telescope. A Kolmogorov phase screen was placed in the aperture of the
telescope and a uniform distribution of rays in the pupil was
assumed. The $x$ and $y$ derivatives of the phase screen were used to
determine the atmospherically induced deviations of the rays as they
propagated toward the focus of the telescope. At the focus, the
incoming rays are mapped onto a fiducial 2D detector grid which
determines the intensity distribution in the focal plane.

\section{Calculating Ellipticity}\label{calc}

\begin{figure}[t]
\plotone{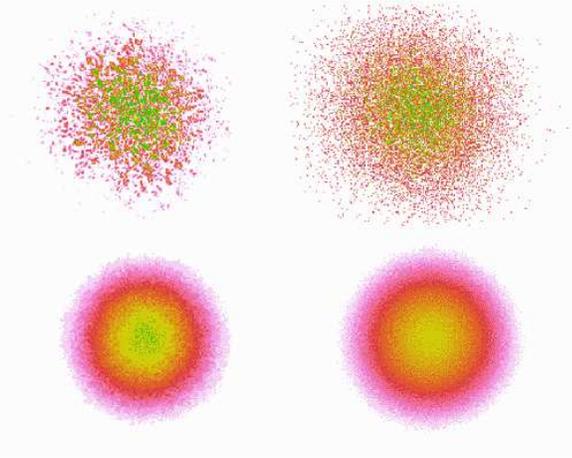}
\caption{Representative point-spread functions, using the Fourier
method (left two images), and the ray-tracing method (right two
images). The top images are for an instantaneous realization of the
atmosphere, whereas the bottom images are the means for 256 such
instances. Notice that the top left Fourier image displays a prominent
speckle pattern due to interference. This pattern gets washed out over
time.}
\label{psfGal}
\end{figure}

We calculate the ellipticity of an object in the pupil plane as
follows. Assume we have an image for which the pixel coordinates are
given by $x = 1 .. N, y = 1 .. N$; the intensity in each pixel is
given by $I(x,y)$; the seeing is given by $\sigma$ (FWHM=2.355$\sigma$
for a Gaussian), and the central point source is located at $(x_c,
y_c)$. We then define Gaussian weights as follows:

\begin{equation}
w(x,y) = \frac{1}{\sqrt{(2\pi\sigma^2)}}
e^{-\frac{1}{2\sigma^2}(x-x_c)^2} . \frac{1}{\sqrt{(2\pi\sigma^2)}}
e^{-\frac{1}{2\sigma^2}(y-y_c)^2}
\end{equation}

\noindent and the Gaussian weighted moments: 
\small
\begin{eqnarray}
S_{w} = \sum_{x,y=1}^N (I(x,y) w(x,y)), \, S_{xx} = \sum_{x,y=1}^N (x^2 I(x,y) w(x,y))\nonumber\\ 
S_{x} = \sum_{x,y=1}^N (x I(x,y) w(x,y)), \,  S_{yy} = \sum_{x,y=1}^N (y^2 I(x,y) w(x,y))\nonumber\\
S_{y} = \sum_{x,y=1}^N (y I(x,y) w(x,y)), \, S_{xy} = \sum_{x,y=1}^N (x y I(x,y) w(x,y))\nonumber\\
\end{eqnarray}
\normalsize

\noindent This allows us to define the following quantities (all
weighted by $I(x,y)w(x,y)$):

\begin{eqnarray}
x_c = S_{x} / S_{w}, \quad y_c = S_{y} / S_{w}\nonumber\\
\left(\begin{array}{c}rxx \\ryy\\rxy \end{array}\right) = \left(\frac{1}{S_{w}S_{w}}\right) \left(\begin{array}{c}S_{xx}S_{w} - S_{x}S_{x} \\S_{yy}S_{w} - S_{y}S_{y}\\ S_{xy}S_{w} - S_{x}S_{y}\end{array} \right)
\end{eqnarray}

\noindent From these, one can calculate the image Gaussian scale
length $\sigma$ and the ellipticity $\epsilon$:

\begin{eqnarray}
\sigma = \sqrt{rxx+ryy}\nonumber\\
\epsilon = \left( \frac{((rxx-ryy)^2+(2rxy)^2}{(rxx+ryy)^2} \right)^{1/2}\nonumber\\
\epsilon_1 = \frac{rxx-ryy}{rxx+ryy}, \quad \epsilon_2 = \frac{2rxy}{rxx+ryy}
\end{eqnarray}

\noindent This is the method used by \citet[][KSB]{kaiser95} - see
also \citet{heymans06} for an overview of the different weak lensing
pipelines.  The weights in eqn.~1 are dependent on initial values of
$\sigma$, and the source centroid $(x_c, y_c)$. However, one can
iterate from initial guesses for the unknowns.  This algorithm
converges quickly (typically within a few steps), and yields values
for $\sigma$ and the source position that are readily verifiable. We
terminate the iterations when the changes in $\sigma$ are less than
0.01 pixel.

\section{Results}\label{results}

\begin{figure*}[t]
\epsscale{1.8}
\plottwo{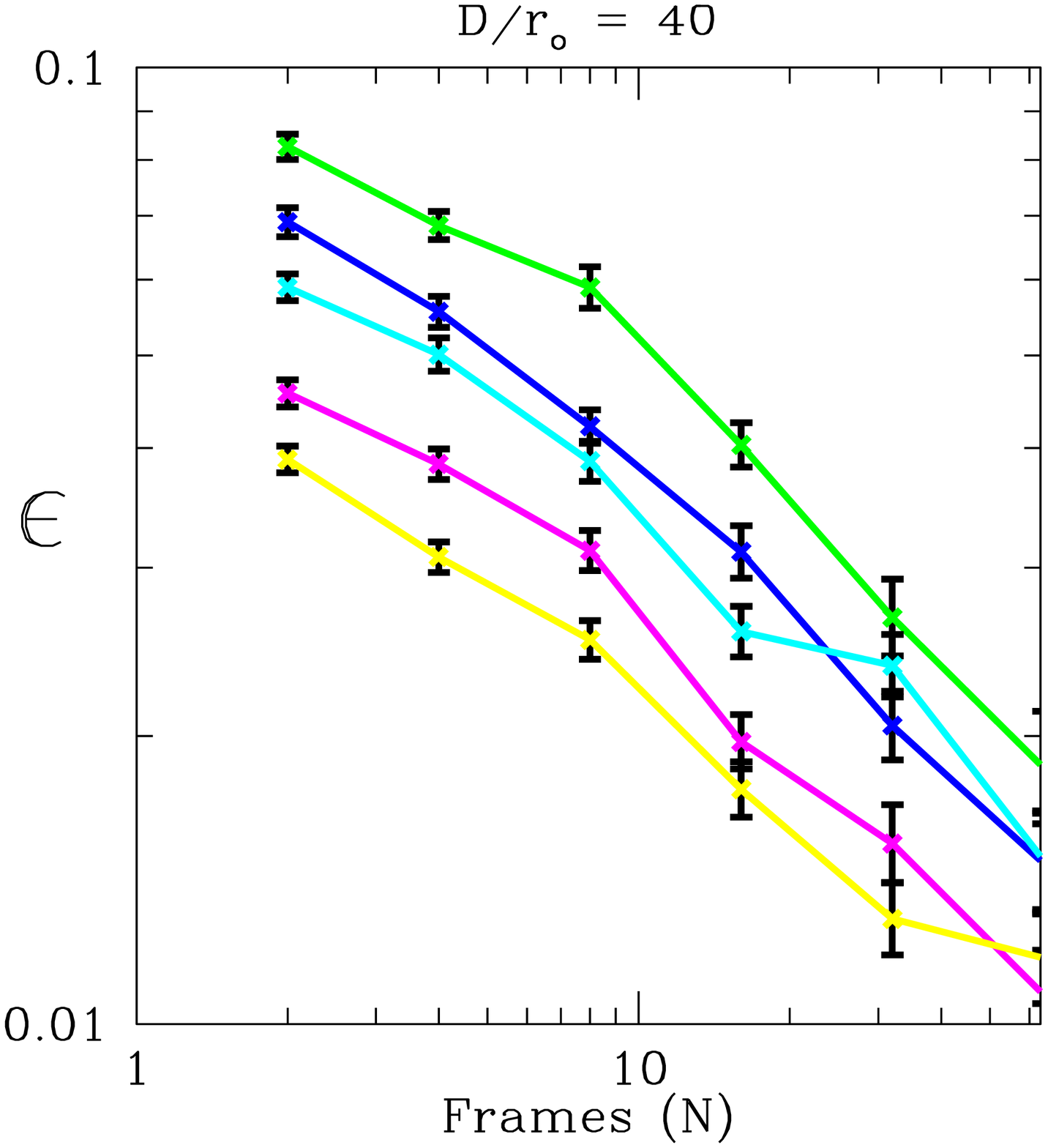}{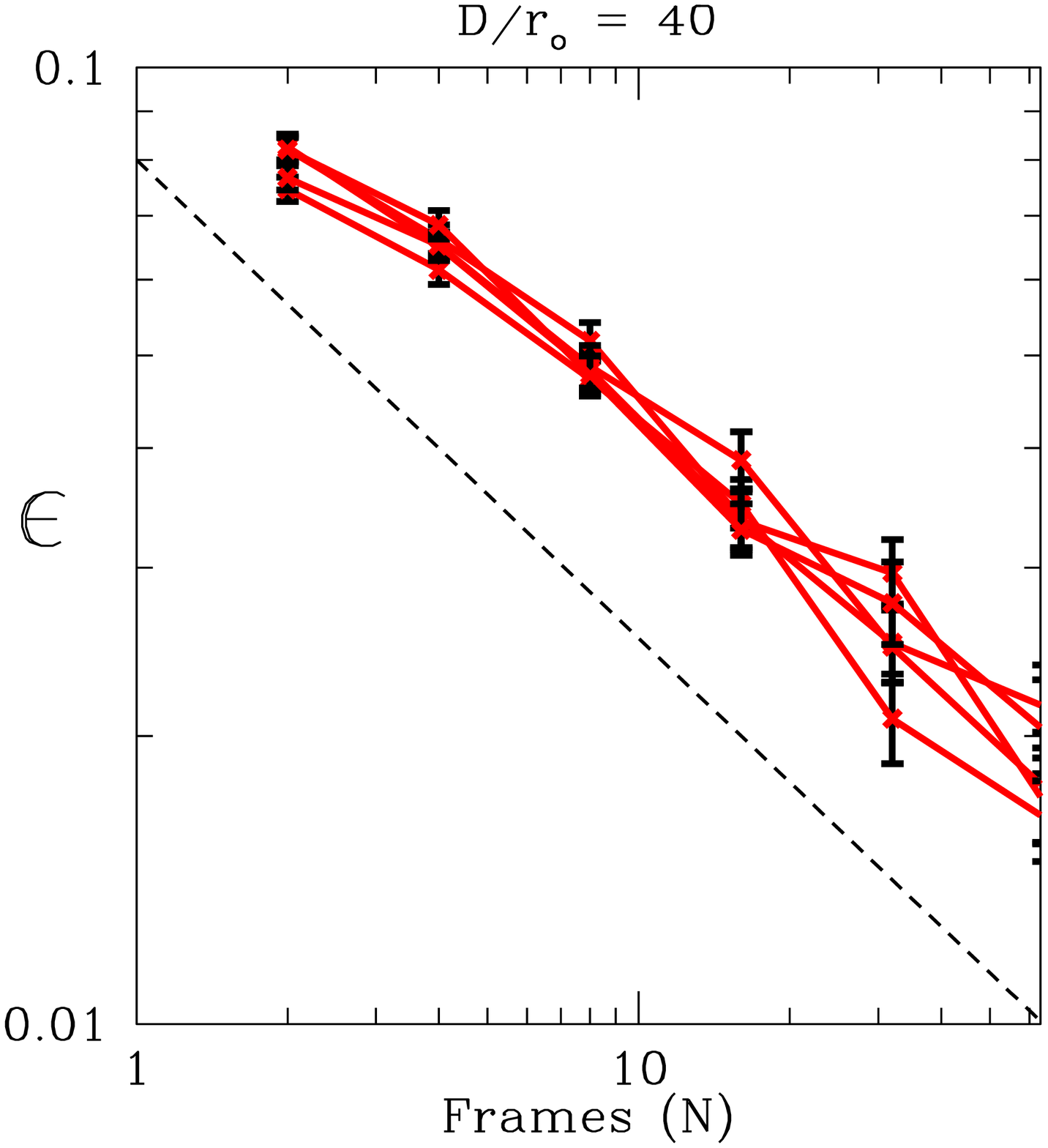}
\caption{Ellipticity as function of the number of independent phase
screens and sampling rates for D / \ro = 40 (\ro = 21cm). The left
panel shows the results for ray-tracing, with the pupil-plane sampling
rates color-coded as: \ro / 2 green, \ro / 4 blue, \ro / 5 cyan, \ro /
8 purple, and \ro / 16 yellow. The panel on the right shows the same
results (except \ro / 5) for the Fourier method. The latter method
clearly illustrates the expected behavior: ellipticity should be
independent of pupil plane sampling rates beyond Nyquist rates. The
progressive lowering of the curves for higher samping rates in the
left panel, therefore, is unphysical. The dashed line in the right
panel shows a $1/\sqrt{N}$ decline.}
\label{samp}
\end{figure*}

Most of the figures in this paper show the ellipticity behavior as a
function of the number $N$ of independent phase screens. A single
phase screen ($N=1$) therefore represents the {\it instantaneous}
ellipticity of a particular representation of the atmosphere (see
Fig.~\ref{psfGal}). In all of our subsequent discussions we assume
that the telescope is perfect, i.e., it does not induce image
aberrations.

Whether we apply the RTM or the FTM approach, we first create stacks
of 500 completely uncorrelated instances of the atmosphere (actually
phase-screens in the pupil plane). The resulting focal plane images
are then either ray-traced or calculated using an appropriate Fourier
transform. After this, we randomly select $N$ frames out of the 500,
which are then stacked, averaged, and have their ellipticity
calculated. We repeat this $1024/N$ times. Since there are not
$1024/N$ fully independent stacks present (for $N>2$), some smoothing
occurs, especially for the larger $N$ stacks. The figures show the
mean ellipticity for these $1024/N$ stacks, and the error-bars on the
means are approximated by the rms of the distribution divided by the
square root of the number of stacks ($1024/N$).

\subsection{Constant ratio D / \ro, varying sampling in pupil plane}\label{sampling}

The value of D / \ro has been fixed at 40 (\ro = 21 cm) for an assumed
8.4 m pupil diameter with a central obscuration\footnote{These are the
current parameters for the LSST design} of 5.4 m, while the (phase)
sampling rate in the pupil plane increases from \ro / 2 to \ro /
16. The ellipticity $\epsilon$ as function of the number of
independent phase screens $N$ is shown in Fig.~\ref{samp}. The left
panel illustrates the results for the RTM method. While each
individual sampling rate falls off as $1/\sqrt{N}$, they are offset in
ellipticity as the sampling rate increases. An increase from \ro / 2
to \ro / 16 more than halves the measured ellipticity for a given
number of independent phase screens. This is clearly not physical.

The following example may illuminate this behavior. Assume, for
instance, a simple one-dimensional cosine phase wave with frequency
$k$. The Fourier transform of this function produces two delta
functions located at $\pm k$ on the x-axis. However, since ray-tracing
uses the derivatives of the phase to calculate where the rays will go,
it produces a spread of points due to the range in derivatives. An
increase in the sampling rate will increase the likelihood high angle
rays will be modeled.  Furthermore, the derivative range becomes
larger for higher spatial frequencies (it goes from $-k$ to $+k$)
resulting in progressively more aberrant rays, regardless of the
sampling rate. While perhaps an extreme example, it does underline the
fact that 2D ray-tracing will produce a more spread-out image due to
rays being deflected into unphysical angles. This broadening of the
image then results in a lowered ellipticity as it decreases the local
asymmetry (remember that the ellipticity contribution of a point is
weighted by its distance, see Eqn. 1).

The FTM, on the other hand, does display the correct behavior (for the
exact same sets of phase screens): the ellipticity is independent of
the sampling rate (provided it is at least \ro / 2).  Except for the
sampling dependency, both methods exhibit the following
characteristics. First, ellipticities decrease linearly (in log-log)
as the number of frames increases. The slope is consistent with a $1 /
\sqrt{N}$ decline ($\alpha=-0.5$), as indicated by the dashed line in
the right panel. And second, ellipticities of individual speckle
images ($N=1$) are $\sim$9\% for D / \ro = 40 (\ro = 21 cm).

It should also be noted that there are $\sim20 - 40$ independent
instances of the atmosphere for an 8.4 meter telescope aperture with
wind-speeds of $\sim10-20$ m/s (typical, turbulence-weighted values
for many astronomical sites) and a 15 second exposure. Therefore,
these simulations predict that the raw ellipticity of a 15-second
exposure for a point source image through an 8.4 meter telescope with
D / \ro = 40 is $\sim2$\%. We explore the effects of wind in more
detail in \S~\ref{wind}.

\subsection{Binning in the image plane}\label{binning}

Obviously, no astronomical instrument designed for seeing-limited
observations will sample the PSF at the Nyquist interval for the
telescope diffraction pattern (we measure $\sigma\sim 60$ pixels), so
the next step is to see what happens to the ellipticities if one
progressively rebins the images of Fig.~\ref{psfGal}. The results are
listed in Table~1.  The ellipticities are for a stack of 32 random
frames with D / \ro = 40. This stack gets increasingly rebinned down
to scales where the PSF is barely resolved ($\sigma \sim 1$).

Given that the values of $\epsilon$ do not change significantly over a
large range of binning, it is clear that the ellipticity measurements
are robust and do not depend on the pixel scale. For comparison, the
rms spread in the value of $\epsilon$ for distinct random stacks of 32
images is about 60\% of the mean value of $\epsilon$, whereas the
listed relative range in column 3 is only $\sim 5\%$ under rebinning.

\begin{deluxetable}{cccc}
\tablewidth{9cm}
\tablenum{1}
\label{bintab}
\tablecaption{Pixel binning effects on ellipticity}
\tablehead{
  \colhead{Binning $n$} &
  \colhead{$\sigma_n$ [pix]} &
  \colhead{$\epsilon_n$} &
  \colhead{Ratio $\epsilon_n / \epsilon_1$ [\%]}
}
\startdata
1 & 61.82 & 0.02505 &  \\
2 & 30.92 & 0.02507 & 100.1 \\
4 & 15.47 & 0.02504 & 100.0 \\
8 &  7.74 & 0.02513 & 100.3 \\
16 & 3.89 & 0.02492 &  99.5 \\
32 & 1.97 & 0.02393 &  95.5 \\
64 & 1.04 & 0.02443 &  97.5 \\
\enddata
\tablecomments{The ellipticity is calculated on a random stack of 32
speckle patterns, with a sampling of \ro / 2, and a ratio of D / \ro =
40 (\ro = 21 cm). There is no significant dependency on pixel
size. Note that $\sigma\sim 1$ to $2$ are typical astronomical seeing
disk sampling ratios.}
\end{deluxetable}

\subsection{Atmospheric model with wind}\label{wind}

\begin{figure}[t]
\epsscale{1.0}
\plotone{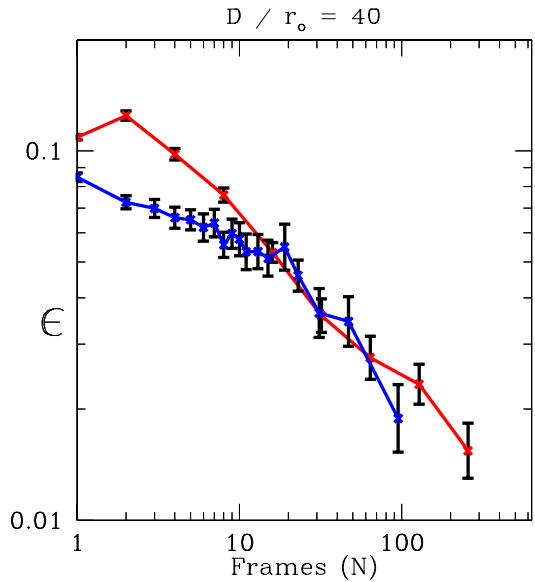}
\caption{Effect of wind on ellipticity behavior for D / \ro = 40 (\ro
= 21 cm). The blue dashed curve is for phase screens which are
translated across the aperture, and the red curve is for independent
instances of the atmosphere. Both these curves are calculated with the
Fourier method. The wind-speed $v$ is needed to convert the number of
frames $N$ into an elapsed time $t$ ($= ND / v$, with $D$ = 8.4 m, and
$v$ in units of m s$^{-1}$). Therefore, a typical 15 s LSST exposure,
with a wind-speed of 10 m s$^{-1}$, contains 18 frames. }
\label{windplot}
\end{figure}

So far we have only considered the behavior of ellipticity as function
of the number of uncorrelated instances of the atmosphere. As
described in \S~\ref{sampling}, the evolution of the PSF with
increasing exposure time can be estimated using the results from these
uncorrelated screens by associating each screen with a unit of time
equal to the aperture diameter divided by the wind speed, i.e., the
time it would take for the wind to translate a screen completely out
of the aperture.  However, a more realistic treatment of the PSF
evolution involves a more continuous translation of a Kolmogorov phase
screen across the telescope aperture.

For this purpose, we generated 3 large phase screens which contain 95
aperture-clearings each. The translation offset\footnote{This offset
should not be confused with the wind-speed $v$. All we want to make
sure is that we have enough numerical resolution (hence the 20 steps)
as the atmosphere translates across the aperture. How long it takes
for the atmosphere to clear an aperture does not matter for this
calculation. See also the caption to Fig.~\ref{windplot}.} is such
that an aperture-clearing takes 20 steps; from each phase screen we
therefore generate 1900 pupil images using the FTM. The ellipticity is
then calculated on combined stacks of 20 pupil images, yielding one
value per aperture-clearing. The results are plotted in
Fig.~\ref{windplot}. The blue dashed line shows the mean ellipticity
values for all $3\times95 / N$ independent aperture-clearings
(``frames''), with as error-bars the error in this mean. At $N=5$, for
instance, we calculated the mean of all uncorrelated instances of
$5\times20$ {\it consecutive} pupil images. On the other hand, the red
curve shows the ellipticity behavior of the sum of $N$ uncorrelated
instances of the atmosphere, using the same pupil plane sampling and
value of D / \ro = 40 (\ro = 21 cm).

A few things stand out. Below about 10 frames or so, the curves behave
differently. In case of the red line, the ellipticity jump from $N=1$
to $N=2$ is due to the fact that there is no ``image motion'' in the
$N=1$ case, whereas for $N=2$, two pupil images have been combined
with different PSF centroids. This raises the ellipticity beyond what
is there in a single PSF (one would need to shift-and-add to remove
this effect). It subsequently takes a few more co-added frames for
this centroid-offset effect to cancel out (on average the PSF has to
align with the optical axis since we put the source there).

The blue line does not suffer from this centroid-offset problem since
the phase-screens are continuous (only 1/20$^{\rm th}$ gets shifted
out between pupil images) and the PSF centroid cannot move around
discontinuous as a consequence. However, we do see another effect
present in the blue curve. Because the Kolmogorov phase screen will
generally have low-spatial-frequency correlations that are larger than
the telescope aperture, the ellipticity of the PSF is expected to
decrease more slowly with increasing $N$ than for the discontinuous
model using multiple independent phase screens. After some time (or
equivalently, for larger values of $N$), these low-spatial-frequency
correlations disappear and the slope of the blue curve steepens to
that of the red curve. This appears to be happening between $N=10$ and
$N=20$.

The main result of this exercise, however, is the confirmation that
our method of using uncorrelated instances timed at a rate equal to an
aperture clearing time is a valid approximation for the ellipticity
behavior in a statistical sense, as long as we are in the
long-exposure, $N > 20$ domain.

\subsection{Varying \ro}
\label{seeingEff}

\begin{figure}[t]
\epsscale{1.1}
\plotone{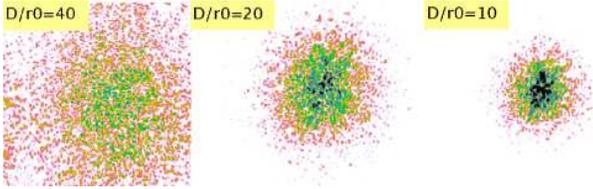}
\caption{Speckle patterns for individual phase screens at various
ratios of D / \ro (\ro = 21, 42, and 84 cm, respectively). Note that
with increasing values of \ro (left to right), the number of speckles
decreases, while their intensities go up. Also, the size of the
pattern decreases with increasing \ro.}
\label{speckle}
\end{figure}

We also investigated the effect of increasing the value of \ro.  The
expectation is that for larger values of \ro (i.e., better seeing
conditions) the number of speckles goes down (no atmosphere = no
speckles, diffraction pattern only) while their individual brightness
goes up (due to the conservation of flux). This is easily verified in
the individual speckle patterns (see Fig.~\ref{speckle}). It is not
clear, however, what the behavior of the RTM method with respect to
the FTM will be. In \S~\ref{sampling}, we noticed that the RTM method
significantly underestimates the actual ellipticity depending on the
sampling rate. If the relative offsets are constant then one might be
able to come up with a particular sampling rate for the ray-tracing
case that best matches the actual ellipticities (for the D / \ro = 40
case the best matching sampling looks to be about \ro / 3, see
Fig.~\ref{samp}).

\begin{figure}[t]
\epsscale{1.0}
\plotone{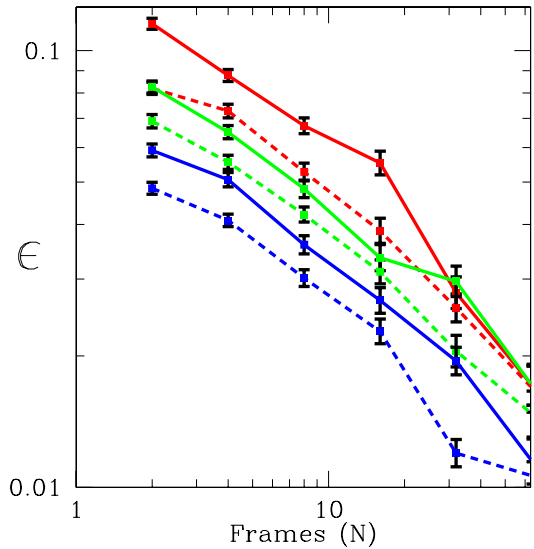}
\caption{Measured ellipticities $\epsilon$ (see eqn. 4) for different
values of \ro, with values from top to bottom of 42 (red, D / \ro
=20), 21 (green, D / \ro = 40), and 10.5 cm (blue, D / \ro = 80),
respectively. The solid line curves have been calculated using the
Fourier method, and have a fixed pupil plane sampling rate of \ro /
4. The dashed lines are calculated using ray-tracing, and are
color-coded and sampled similarly. The ellipticity for a given number
of independent phase screens depends on the size of \ro: large values
of \ro have larger ellipticities. Also note that low number statistics
are affecting the data-points toward large frame counts causing the
curves to cross eachother.}
\label{seeing}
\end{figure}

The results are presented in Fig.~\ref{seeing}, with the solid lines
representing the FTM results, and the dashed lines are for the RTM
method. If we focus on the solid lines first, it is clear that the
ellipticities increase with increasing values of \ro at the same
number of stacked frames.  This can be qualitatively understood in
terms of the decreasing number of speckles distributing themselves in
a progressively less circular pattern due to the smaller number
statistics.

Based on this plot, it is also apparent that, even though the RTM
underestimates the ellipticity compared to the FTM, it does so more or
less independently of the value of \ro (which is a proxy for
seeing). This might open up the possibility that one either selects a
computationally efficient RTM sampling rate (say, \ro / 2) and apply
an appropriate (fixed) correction factor to the ellipticity results,
or adjust the sampling rate such that the RTM and FTM results are in
good agreement ($\sim$\ro / 3). Another approach that is under
investigation (G. Jernigan, private communication) is to roll off the
atmospheric power spectra at high spatial frequencies.  Further study
is needed to assess the accuracy of any of these approaches. For
instance, the numerical correction factors derived from the results
shown in Fig. 5, for the particular sampling ratio of \ro / 4, are
1.32, 1.15, and 1.21, for \ro = 42 cm, \ro = 21 cm, and \ro = 10.5 cm,
respectively. Whether this variation is due solely to the statistical
errors in our modeling, or includes a systematic dependence on \ro is
not known. Clearly if one requires accurate modeled ellipticities,
then the computationally more expensive FTM is currently preferred.

\section{Conclusions}\label{conclusion}

Based on this analysis, we reach the following conclusions:

\begin{enumerate}
\item{Instantaneous speckle patterns have ellipticities of $\sim$10\%
for D / \ro = 40 (\ro = 21 cm).}
\item{Co-adding multiple patterns results in a linearly decreasing
ellipticity (on a log-log plot), consistent with a $\sqrt{N}$ slope of
$\alpha=-0.5$. }
\item{Modeling phase screen transport across the aperture (i.e., wind)
does not change the ellipticity results obtained from adding
uncorrelated phase screens in the limit of long exposures ($N>20$
aperture clearings).}
\item{Ellipticity values are robust over a large range of pixel
binning.}
\item{We expect no ellipticity dependency on sampling density in the
pupil plane (above sampling of \ro / 2). This is confirmed for the
Fourier method, but not for ray-tracing. The latter has a strong
dependency on sampling rate, in the sense that the higher the sampling
rate, the lower the resulting ellipticity. This can be understood as
the result of a breakdown in the geometric approximation for high
spatial frequency aberrations.}
\item{Ellipticities grow (for a given $N$) as the value of \ro
increases. However, since the average {\it size} of the PSF goes down
as \ro increases (see Fig.~\ref{speckle}), the ability to measure
precise ellipticities actually improves (for a given resolved object).}
\end{enumerate}

\noindent In summary, the effects of interference must be included in
order to comprehensively model point-source ellipticities induced by
the atmosphere.  Therefore, care has to be taken that the geometric
optics approximation to image formation by the telescope (i.e., ray
tracing) produces the same modeling results, as this is not true in
general.

\acknowledgments

The authors like to thank Don Phillion for generating the large phase
screens of \S~\ref{wind}, and the anonymous referee for helpful
comments.  This work was performed under the auspices of the
U.S. Department of Energy, National Nuclear Security Administration by
the University of California, Lawrence Livermore National Laboratory
under contract No.  W-7405-Eng-48.

\end{document}